\documentclass[11pt,letterpaper]{article}
\usepackage[margin=1in]{geometry}
\usepackage[T1]{fontenc}
\usepackage{lmodern}
\usepackage{amsmath,amssymb,amsthm}
\usepackage{booktabs,tabularx}
\usepackage{microtype}
\usepackage{xcolor}
\usepackage[section]{placeins}
\usepackage{enumitem}
\usepackage{needspace}
\usepackage{xurl}
\usepackage{hyperref}
\hypersetup{colorlinks=true,linkcolor=blue!50!black,citecolor=blue!50!black,
 urlcolor=blue!50!black,
 pdftitle={Accuracy Is Not Risk},
 pdfauthor={Jeffrey Humpherys and Makoto Jones},
 pdfsubject={Uncertainty propagation, label compression, and downstream risk in medical AI}}
\newcommand{\given}{\,|\,}
\newcommand{\argmax}{\operatorname*{arg\,max}}
\newcommand{\Rsoft}{R_{\mathrm{soft}}}
\newcommand{\Rhard}{R_{\mathrm{hard}}}
\newcommand{\indep}{\perp\!\!\!\perp}
\newcommand{\E}{\mathbb{E}}
\newcommand{\Var}{\operatorname{Var}}
\newcommand{\TV}{\operatorname{TV}}
\newtheorem{proposition}{Proposition}[section]
\newtheorem{corollary}[proposition]{Corollary}
\theoremstyle{remark}

\title{Accuracy Is Not Risk\\[4pt]
{\large Why Medical AI Should Not Report Only Its Most Likely Label}}
\author{\begin{tabular}{c}
Jeffrey Humpherys, PhD\\
\small Department of Mathematics and Statistics\\
\small Missouri University of Science and Technology\\
\small\texttt{jhumpherys@mst.edu}
\end{tabular}
\and
\begin{tabular}{c}
Makoto Jones, MD, MS\\
\small Department of Internal Medicine, Division of Epidemiology\\
\small University of Utah School of Medicine\\
\small\texttt{makoto.jones@hsc.utah.edu}
\end{tabular}}
\date{October 23, 2026}

\begin{document}
\maketitle

\begin{abstract}
\noindent\textbf{Background.} Many medical AI models are trained to reproduce a single reference label and hand the most likely label to the next step in a clinical workflow. A classifier can select that label correctly while providing an inadequate summary for estimating downstream risk.
 
\smallskip
\noindent\textbf{Methods.} We use Jeffrey's rule of conditioning to specify when probabilities over intermediate states, such as diagnoses or imaging categories, can be carried forward to an outcome risk. We separate two errors of reporting only the most likely label, discarding the probability of the other states and applying the risk of patients who have the leading diagnosis to a patient who only probably has it. We also give conditions under which the label and the full distribution lead to the same decision, and we propose guardrails for device developers, regulatory reviewers, and health insurers.

\smallskip
\noindent\textbf{Results.} In a hypothetical chest-pain differential, reporting only the benign leading diagnosis greatly understates the patient's risk of death or cardiac arrest, so a patient who warrants evaluation could be discharged or denied imaging. In a hypothetical BI-RADS example, the most likely category is ``probably benign,'' yet the propagated malignancy risk exceeds the ceiling that defines that category.
 
\smallskip
\noindent\textbf{Conclusions.} A predicted label is not an observed state, and a calibrated classifier is not automatically a valid outcome model. Clinical and insurance coverage decisions should rest on the patient's outcome risk, judged against a threshold justified by the consequences of acting and not acting. Developers should state what each model output means and check that every step between the model and the decision preserves the patient's risk.
\end{abstract}
 
\noindent\begin{minipage}{\textwidth}
\section*{Highlights}
\begin{itemize}[leftmargin=1.4em,itemsep=2pt,topsep=2pt]
\item Training to a label and treating the predicted label as truth can cause distinct, quantifiable errors.
\item A calibrated classifier does not guarantee a correct risk for the patient.
\item Insurance coverage rules keyed to a diagnosis code can deny care to patients whose risk is high.
\item Six guardrails tell developers, reviewers, and payers what to report and what to check so that risk is not lost between the model and the decision.
\end{itemize}
\end{minipage}
 
\section{Introduction}
\label{sec:introduction}
 
A patient arrives in the emergency department with chest discomfort. The history is ambiguous and the examination is unremarkable. An AI tool trained on many thousands of similar presentations reports that the most likely explanation is gastroesophageal reflux. Suppose the tool is right, in the sense that reflux is indeed the single most likely diagnosis, with probability just over one half. Suppose also that the tool's internal calculation assigned nearly a one-in-three chance that the cause is instead acute coronary syndrome, pulmonary embolism, or aortic dissection. No experienced clinician would discharge this patient on the strength of the leading diagnosis. The leading diagnosis is benign, but the patient is not low risk.

Nothing about this reasoning is new to physicians. The differential diagnosis exists because the most likely explanation is not the only one that matters, and the threshold approach to clinical decision making has long made action depend on the probability of disease together with the consequences of acting or not acting, rather than on the identity of the leading diagnosis \cite{pauker1980}. What is new is the ease with which an AI system can undo this reasoning without anyone noticing. Many medical AI models are trained to reproduce a single reference label per case, are judged by how often their top label matches it, and hand that top label to the next step in the clinical workflow. At that moment the one-in-three chance of a serious cause disappears. A downstream calculation, a triage rule, an order set, a health insurance coverage decision, or a discharge summary then treats ``reflux'' as the patient's condition rather than as the model's best guess. We call the replacement of a probability distribution by its most likely label \emph{hardening} the prediction. The same failure appears in imaging, where a tool that reports ``probably benign'' with 70\% confidence may be describing a lesion whose propagated malignancy risk exceeds the 2\% ceiling that defines the category.

Hardening causes two distinct errors. The first is that it discards information. A label says nothing about how confident the model was or where the remaining probability lay, so two patients with the same leading diagnosis and the same confidence can face very different risks, if the rest of the probability sits on a second benign diagnosis for one and on a life-threatening condition for the other. The second is that it attaches the wrong risk to the label. Downstream users take the risk of patients who actually have a diagnosis and apply it to patients the model has only predicted to have it. Patients labeled as having reflux can die more often than patients who have reflux, because the labeled group includes the missed dissections. A predicted label is not an observed state, and this paper is organized around keeping the two errors apart.

There is a classical way to carry uncertainty forward instead of discarding it. If the model's probabilities for the competing states are trustworthy, and if the risk of the outcome within each state does not depend on the particular patient beyond membership in that state, then the patient's outcome risk is the probability-weighted average of the state risks. This is Jeffrey's rule of conditioning \cite{jeffrey1983,shafer1981,diaconis1982}. In the chest-discomfort example, the leading diagnosis alone suggests a patient at almost no risk, while the weighted average reveals a substantial risk of death or cardiac arrest. We show that the second condition can fail, and that when it does, propagation can be worse than a well-chosen label even when the model is well calibrated and its labels are accurate.

Each piece of this argument has precedents. Jeffrey's rule is a standard way to enter uncertain evidence into Bayesian networks \cite{valtorta2002,pan2006,jacobs2019}, and it has been used with medical and forensic classifiers \cite{corradi2013,rabaey2026}. Machine-learning research has shown that a model can pick the right label while reporting probabilities that are too confident \cite{guo2017}, that probabilities can be reliable on average without being right for each patient \cite{vaicenavicius2019}, and that the reference labels used to train medical models are themselves uncertain \cite{mimori2021}. Decision curve analysis and cost-sensitive learning build on the threshold approach \cite{vickers2006,elkan2001}, and statisticians have studied the bias that arises when estimated classes are treated as known \cite{bray2015}. Several authors have called for medical AI to communicate its uncertainty \cite{begoli2019,kompa2021,harish2021}, and the FDA's regulatory-science program studies how to quantify it \cite{fda_uq}.

Our contribution is to assemble these ideas into a task-specific account of a two-stage prediction failure and to make it usable by the people who build, review, and pay for medical AI. Section~\ref{sec:examples} presents two hypothetical clinical examples. Section~\ref{sec:hardening} states what hardening changes, separates information loss from the misattributed risk, gives conditions under which a hard label and the propagated risk lead to the same decision, and shows how a change in case mix alters both the risk and the label. Section~\ref{sec:payers} applies the analysis to prior authorization in health insurance, and Section~\ref{sec:guardrails} collects the consequences in six guardrails. The results are elementary applications of probability and decision theory, not a new updating rule. Proofs, the complete notation, and extensions to truncated outputs and model error are in the Appendix.

\section{Two Clinical Examples}
\label{sec:examples}

Both examples share one probabilistic setup. Consider a population of patient encounters in a specified setting, with probability $P$ describing how encounters in that population vary. For a randomly chosen encounter, the random variable $X$ is the information supplied to the model, with observed value $x$ for a particular patient. The random variable $E$ is the patient's intermediate state, such as the primary diagnosis or the imaging category, taking one of $n$ mutually exclusive values $1,\ldots,n$. The event $A$ is the outcome of interest, such as death or cardiac arrest within 30 days, or malignancy. The quantity a downstream decision requires is the patient's outcome risk
\[
\eta(x)=P(A\given X=x),
\]
which conditions on what has actually been observed about the patient.

Two inputs are available for estimating $\eta(x)$. The classifier is a fixed function $q$ that maps patient information $x$ to reported probability weights $q(x) = (q_1(x),\ldots,q_n(x))$, which are intended to equal the state probabilities $p_i(x)=P(E=i\given X=x)$. A risk table supplies
\[
r_i=P(A\given E=i),\quad i=1,2,\ldots,n,
\]
the outcome probabilities among patients who are actually in each state $i$. The state probabilities $p_i(x)$ condition on each patient's information and vary from patient to patient, whereas the table entry $r_i$ conditions only on the state and is the same for every patient in that state.

Total probability, applied conditionally on $X=x$, gives the exact identity
\begin{equation}
\label{eq:total-prob}
\eta(x)=\sum_{i=1}^n P(A\given E=i,X=x)\,P(E=i\given X=x).
\end{equation}
The \emph{propagated risk} carries the classifier's probabilities through the risk table,
\begin{equation}
\Rsoft(x)=\sum_{i=1}^n q_i(x)\,r_i .
\label{eq:soft-risk}
\end{equation}
The propagated risk equals the patient's risk $\eta(x)$ in \eqref{eq:total-prob} under two assumptions. The first is that the classifier is correct, so that $q_i(x)=p_i(x)$. The second is that once the state is known, the patient information says nothing more about the outcome, so that $P(A\given E=i,X=x)=r_i$. Substituting both into \eqref{eq:total-prob} gives $\Rsoft(x)=\eta(x)$, and Section~\ref{sec:hardening} examines what happens when either assumption fails. Hardening takes a shortcut. It selects the most likely state $c(x)=\argmax_i q_i(x)$ and reports that state's risk, the \emph{modal-state lookup}
\[
\Rhard(x)=r_{c(x)}=P(A\given E=c(x)).
\]
This is the risk for patients known to be in state $c(x)$. Nothing about this patient has established that state, so the lookup conditions on the wrong event, the unobserved $E=c(x)$ rather than the observed $X=x$. All numerical values below are hypothetical, chosen to illustrate the calculations rather than to represent clinical event rates, and Table~\ref{tab:notation} in Appendix~\ref{app:notation} collects the notation.

\subsection{AI-generated differential diagnosis}
\label{sec:ddx}

Let us return to the chest-discomfort patient, with the five possible primary etiologies in Table~\ref{tab:ddx-risks}. A practical model would need more causes and would have to allow for more than one, but for illustrative purposes we assume the states are mutually exclusive and collectively exhaustive. The state $E$ is the primary etiology, and the outcome $A$ is death or cardiac arrest within 30 days. The risks $r_i = P(A\given E=i)$ are event rates among patients with each etiology who receive usual care, so they apply only as long as the AI output does not change that care \cite{vangeloven2020}.

\begin{table}[htbp]
\centering
\begin{tabular}{lcc}
\toprule
Primary etiology $E=i$ & $q_i(x)$ & Illustrative $r_i$\\
\midrule
Gastroesophageal disease (GD) & $0.52$ & $0.001$\\
Musculoskeletal pain (MP) & $0.18$ & $0.001$\\
Acute coronary syndrome (ACS) & $0.18$ & $0.15$\\
Pulmonary embolism (PE) & $0.08$ & $0.10$\\
Aortic dissection (AD) & $0.04$ & $0.50$\\
\bottomrule
\end{tabular}
\caption{Hypothetical model output $q_i(x)$ for one patient, assumed equal to $p_i(x)= P(E=i\given X=x)$, and illustrative 30-day event rates $r_i=P(A\given E=i)$ under one fixed reference care policy.}
\label{tab:ddx-risks}
\end{table}

Gastroesophageal disease is the most likely individual diagnosis, with probability 52\%. Yet the three serious alternatives together have probability
\[
P(E\in\{\text{ACS},\text{PE},\text{AD}\}\given X=x)=0.18+0.08+0.04=0.30. 
\]
This is still a state probability, not an outcome risk. It uses only the model output and adds the probabilities of mutually exclusive states, without the risk table.

The adverse event requires the risk table. Propagation gives
\[
\Rsoft=(0.52)(0.001)+(0.18)(0.001)+(0.18)(0.15)+(0.08)(0.10)+(0.04)(0.50)=0.0557,
\]
while the modal-state lookup is $\Rhard=P(A\given E=\text{GD})=0.001$. 

The propagated risk is more than fifty times the modal-state lookup. To see where it comes from, picture 1{,}000 patients who receive this output. About 56 would die or have a cardiac arrest within 30 days, and 55 of them would be among the 300 patients with a serious cause. Aortic dissection, with a probability of only 4\%, accounts for 20 of those events. The modal-state lookup misses all of this, because it applies the event rate of patients who have gastroesophageal disease to a patient who only probably has it. Propagation does not decide what to do, since harms, treatment effects, testing costs, and patient preferences still belong to a decision analysis, but it keeps the probabilities that such an analysis needs.

\subsection{AI-assisted BI-RADS assessment}
\label{sec:birads}

The Breast Imaging Reporting and Data System (BI-RADS) associates assessment categories with malignancy-risk ranges and management recommendations \cite{acr2025}. Category 3 is probably benign, with a malignancy likelihood no greater than about 2\%, while category 4 is suspicious and subdivided into 4A, 4B, and 4C \cite{strigel2017,understandingbirads3}. In mammography, category 3 follows an appropriate diagnostic evaluation rather than substituting for workup of an indeterminate screening finding \cite{understandingbirads3}.

Here the state $E$ is the final assessment that a reference radiologist, following a fixed reading protocol, would assign after that workup. The state is therefore a reader's category rather than a biological fact, and uncertainty in $E$ given the information $X$ available to the AI can reflect omitted information or reader variability. The outcome $A$ is malignancy. We restrict the illustration to $E\in\{3,\mathrm{4A}\}$ and use the hypothetical risks in Table~\ref{tab:birads-risks}, so that $r_3=P(A\given E=3)$ is the malignancy rate among lesions the reference reader would place in category 3. An applied model would require its labels and risk table to use the same reading protocol and population, and outcome ascertainment would have to address differential verification, since surveillance is common for category 3 while tissue diagnosis is generally recommended for category 4 \cite{understandingbirads3,strigel2017}.

\begin{table}[htbp]
\centering
\begin{tabular}{lll}
\toprule
Category $E=i$ & BI-RADS risk interpretation & Illustrative $r_i$\\
\midrule
3 & Probably benign, $\le2\%$ & $0.010$\\
4A & Low suspicion, $>2\%$ to $\le10\%$ & $0.060$\\
\bottomrule
\end{tabular}
\caption{Hypothetical category-conditional malignancy risks, within the category ranges but not clinical estimates or management thresholds.}
\label{tab:birads-risks}
\end{table}

Suppose that for a particular examination the model reports $q_3(x)=0.70$ and $q_{\mathrm{4A}}(x)=0.30$, assumed equal to $P(E=3\given X=x)$ and $P(E=\mathrm{4A}\given X=x)$, respectively. The modal-state lookup is $\Rhard=r_3=0.01$, whereas
\[
\Rsoft=(0.70)(0.01)+(0.30)(0.06)=0.025 .
\]
Among 1{,}000 such examinations, about 700 would be category 3 with 7 cancers, and about 300 would be category 4A with 18 cancers. Category 3 is most likely, yet the propagated risk is 2.5\%. The two percentages condition on different events. The BI-RADS ceiling of 2\% describes $P(A\given E=3)$, the risk among lesions that are category 3, whereas 2.5\% describes $P(A\given X=x)$, the risk for a lesion that is only 70\% likely to be category 3. More generally, since $q_{\mathrm{4A}}(x)=1-q_3(x)$, the propagated risk is $\Rsoft=0.01\,q_3(x)+0.06\,(1-q_3(x))=0.06-0.05\,q_3(x)$. Category 3 is the most likely state whenever $q_3(x)>0.50$, but the propagated risk stays above 2\% until $q_3(x)$ reaches $0.80$.

This calculation does not justify relabeling an examination as BI-RADS 4A. BI-RADS is an imaging assessment, not a numerical bin into which an arbitrary risk estimate can be inserted. The margin above 2\% is also small, and Appendix~\ref{app:sensitivity} shows how modest errors in the risk table could erase it. How the output is displayed matters as well. Dratsch et al.\ found that purported AI BI-RADS suggestions influenced reader performance even when the suggestions were incorrect \cite{dratsch2023}, although that study did not compare categorical with probabilistic displays.

\section{What Hardening Changes}
\label{sec:hardening}

\subsection{When propagation is valid}

The second assumption of Section~\ref{sec:examples} has a name. In Jeffrey's formulation, new information revises the state probabilities to $q$ while the within-state risks are kept, and the assumption that the conditional risks $P(A\given E=i)$ are unchanged by the revision is called \emph{rigidity}. Total probability then gives Jeffrey's rule,
\begin{equation}
P_{\mathrm{new}}(A)=\sum_i q_i\,P(A\given E=i),
\label{eq:jeffrey}
\end{equation}
so the propagated risk \eqref{eq:soft-risk} is a Jeffrey update whenever $q$ is the revised state distribution and the table remains appropriate \cite{diaconis1982,jacobs2019}. In words, Jeffrey's rule says that new information may change how likely each state is, but not how dangerous each state is once a patient is in it. New imaging or laboratory findings can violate that assumption by changing a patient's risk even after the state is fixed.

Calibration does not settle the question. A model is calibrated when, among patients who receive the same output vector, each state occurs with the reported frequency, which is a statement about groups rather than individual patients. The constant predictor that reports population prevalence to everyone is perfectly calibrated even when the patient data are highly informative \cite{vaicenavicius2019}, and calibration of only the winning probability does not imply calibration of the full vector \cite{kull2019}. Appendix~\ref{app:calibration} shows that a calibrated full vector makes $\Rsoft$ the outcome risk among patients who receive the same output, provided the output carries no outcome information beyond the state, whereas equality with the patient's own risk $\eta(x)$ follows from the true posterior and rigidity. Calibration and rigidity are different requirements.

\subsection{The discrepancy and when decisions agree}

As in Section~\ref{sec:examples}, $c$ denotes the most likely state. Proofs of the results in this section are in Appendix~\ref{app:proofs}.

\begin{proposition}[Discrepancy and bound]
\label{prop:gap}
For any probability vector $q$ and risk vector $r$,
\begin{equation}
\Delta:=\Rsoft-\Rhard=\sum_{i\ne c}q_i(r_i-r_c),
\qquad
|\Delta|\le B:=(1-q_c)\max_{i\ne c}|r_i-r_c|.
\label{eq:hardening}
\end{equation}
\end{proposition}

Hardening changes the risk by exactly the discarded probability of each competing state times how much more or less dangerous that state is than the leading one. It can underestimate or overestimate the propagated risk. Following an example of Vaicenavicius et al.\ \cite{vaicenavicius2019}, the vectors $(0.70,0.30,0)$ and $(0.70,0,0.30)$ share the same winning state and confidence, yet with risks $r=(0.01,0.02,0.90)$ their propagated risks are $0.013$ and $0.277$. Confidence in the winner does not determine the downstream effect; where the remaining mass lies does. Choosing the modal state is Bayes-optimal for classifying the state under zero--one loss \cite{elkan2001}, but reusing that choice to estimate an outcome is a separate operation.

For a threshold $\tau$, consider the rule that acts when a risk is at least $\tau$. Such a rule is appropriate only when the outcome and the consequences of the available actions justify the threshold \cite{pauker1980,vickers2006}.

\begin{corollary}[Sufficient condition for decision agreement]
\label{cor:agreement}
If $|\Rhard-\tau|>B$, then $\Rhard$ and $\Rsoft$ lie on the same side of $\tau$ and lead to the same decision.
\end{corollary}

If the hard-label risk is farther from the threshold than hardening could possibly move it, the hard label gives the same decision as propagation. In the BI-RADS example with $q_3(x)=0.70$, $B=0.30(0.06-0.01)=0.015$ exceeds the distance $0.01$ between $\Rhard$ and $2\%$, so the condition fails and the decisions in fact disagree. At $q_3(x)=0.90$, $B=0.005<0.01$ and both calculations fall below $2\%$. The condition is sufficient, not necessary, and agreement between two calculations is not evidence that either is clinically correct.

\subsection{Predicted states are not actual states}
\label{sec:decomposition}

A downstream system that receives only the label $C$ could instead estimate
\[
g(j)=P(A\given C=j),
\]
the risk among patients the model \emph{labels} $j$. This differs from $r_j=P(A\given E=j)$, the risk among patients who \emph{are} in state $j$, and it can be estimated from outcome-labeled validation data without pretending that $C=E$. The difference between $g$ and $r$ is the second error of Section~\ref{sec:introduction}.

\begin{proposition}[Label-compression loss and plug-in error]
\label{prop:brier}
Let $Y$ indicate the outcome, and suppose $S=\Rsoft$ is the outcome probability given the model output, as in Appendix~\ref{app:calibration}. Let $T=r_C$ be the modal-state lookup. Then
\begin{equation}
\E[(Y-T)^2]-\E[(Y-S)^2]
=\underbrace{\E[\Var(S\given C)]}_{\text{label-compression loss}}
+\underbrace{\E[(g(C)-T)^2]}_{\text{plug-in error}}.
\label{eq:brier-decomposition}
\end{equation}
\end{proposition}

The excess squared error of the modal-state lookup splits into the risk variation hidden inside each label, which no label-only system can recover, and the error of attaching the wrong risk to the label, which a label-only system can avoid by estimating $g$. This split parallels the well-known decomposition of the Brier score into reliability and resolution terms \cite{murphy1973}. A fair evaluation of propagation therefore compares it with a properly estimated $g(C)$, not only with the modal-state lookup.

The full vector is not always required. For one fixed outcome and risk table, the single number $\Rsoft$ preserves everything the decision uses. Retaining the vector matters when outcomes, risk tables, or decisions may change. Systems that keep several classes rather than one can still bound the full propagated risk from the retained probabilities and the total omitted probability, provided the omitted mass is reported and not renormalized away. Renormalizing the retained classes to sum to one conditions on those classes and can move the risk in either direction (Appendix~\ref{app:partial}).

\subsection{Three sources of error}
\label{sec:sources}

The discrepancy $\Delta$ compares two calculations, not a calculation with the truth. Recall that $p_i(x)=P(E=i\given X=x)$ is the correct probability of state $i$ for the patient, and let $\delta_i(x)$ be the amount by which the patient's within-state risk differs from the table entry $r_i$. Then
\begin{equation}
\eta(x)-\Rhard=\underbrace{\Delta}_{\text{hardening}}
+\underbrace{\sum_i(p_i-q_i)r_i}_{\text{posterior error}}
+\underbrace{\sum_i p_i\delta_i}_{\text{rigidity failure}} .
\label{eq:total-discrepancy}
\end{equation}
Even a model that reports the true posterior and a table that is accurate on average leave the last term, which can reinforce, cancel, or dominate the hardening term. This is why propagation can be worse than a well-chosen label. Appendix~\ref{app:sensitivity} turns \eqref{eq:total-discrepancy} into an error envelope and a correspondingly widened agreement condition.

A common source of posterior error is deployment in a population where the states occur in different proportions than where the model was developed, because a posterior already incorporates the prevalence of its development population. Suppose the BI-RADS model was developed where categories 3 and $\mathrm{4A}$ occur in proportion 90 to 10 and is used unchanged in a diagnostic clinic where the proportion is 60 to 40, with the same image findings and within-category risks at both sites. The standard correction for new prior probabilities \cite{saerens2002,elkan2001} turns the reported $(0.70,0.30)$ into $(0.28,0.72)$, which raises the propagated risk from 2.5\% to 4.6\% and changes the most likely category from 3 to $\mathrm{4A}$. A downstream rule keyed to the label can therefore behave differently at two sites even when the model and the image are identical. The correction treats the output as likelihood evidence and applies it by Pearl's rule of virtual evidence rather than by Jeffrey's rule \cite{pearl1988,jacobs2019}. Clinicians make the same distinction routinely, combining a test result with a pretest probability but letting a consultant's considered opinion replace their own. An AI output can be either, and its display rarely says which.

\section{Health Insurance Prior Authorization and Coverage Decisions}
\label{sec:payers}

Before some tests and treatments are performed, a health insurer decides whether it will pay for them. This prior-authorization review asks whether the service is medically necessary for the patient, and it is increasingly screened or carried out by automated rules. The physician's office submits a prior-authorization request, which reaches the insurer as diagnosis codes and supporting documentation, so a coverage rule sees only as much of the clinical uncertainty as the request or an AI system passes along. Both errors of hardening described in the Introduction therefore reappear. A rule that reads only the label loses the probability of the other diagnoses, and it applies the risk of the leading diagnosis to a patient who only probably has it. In this setting these errors can determine whether a service is provided at all.

Consider the chest-discomfort patient of Section~\ref{sec:ddx}, for whom the physician requests computed tomographic angiography to exclude the serious causes. Suppose, for illustration, that the insurer covers imaging when the patient's 30-day risk of death or cardiac arrest is at least 2\%. With the full differential of Table~\ref{tab:ddx-risks}, the propagated risk is 5.57\% and the prior-authorization request is approved. Other rules applied to the same patient and the same model output can reach the opposite decision, for the reasons described below.

\paragraph{Reading the label.} The prior-authorization request carries a diagnosis code, which is a hard label. If the code is the leading diagnosis, gastroesophageal disease, a rule keyed to the code sees a benign condition with a risk of 0.1\% and denies. A symptom code such as unspecified chest pain at least avoids asserting a benign diagnosis, but no choice of code conveys the 5.57\% risk on which the decision depends. That number has to travel with the prior-authorization request.

\paragraph{Partial lists.} Suppose the insurer's system receives the model's probabilities only for the diagnoses coded on the prior-authorization request, here gastroesophageal disease and musculoskeletal pain, which together carry 70\% of the probability. If a rule rescales those two to sum to one, it treats the serious causes as impossible and computes a risk of 0.1\%. Which diagnoses are coded then decides the answer. Coding only the benign causes leads to denial, and coding only the serious ones would inflate the risk to about 18\%, yet the party that makes the choice does not bear the clinical consequences, an incentive asymmetry that is a form of moral hazard. The remedy is to report the probability left off the list. Knowing only that 30\% belongs to unlisted diagnoses whose risks range from 10\% to 50\%, a rule can bound the full risk between 3.1\% and 15.1\% (Appendix~\ref{app:partial}) and approve without the full differential.

\paragraph{Requiring the serious diagnosis to be likely.} Even a rule that sees the full differential can deny if it approves imaging only when a serious diagnosis is more likely than not. Here the three serious causes together have a 30\% probability, short of one half, so the rule denies. Yet a test is worth doing whenever the harm it can prevent outweighs its own harms and costs \cite{pauker1980}, and when one of the conditions being excluded leads to death or cardiac arrest in half of those who have it, that point comes at a probability far below one half. The rule fails because it asks whether a serious diagnosis is probable rather than whether the patient's risk is high enough to act on.

A coverage rule consistent with this analysis computes the patient's outcome risk from the full distribution, or bounds it from a partial one with the omitted probability reported, and compares that risk with a threshold justified by the consequences of the available actions. It also fixes the retained diagnoses, the risk table, and the threshold in advance, so that the determination can be audited. For Medicare Advantage, CMS requires medical-necessity determinations to rest on the individual patient's circumstances \cite{cms4201f}, and its subsequent guidance states that algorithms may assist with but may not replace that individualized determination \cite{cms2024faq}. A determination that uses the patient's distribution over diagnoses, rather than the label it was compressed into, is individualized in a precise and auditable sense. The guardrails of Section~\ref{sec:guardrails} apply to coverage algorithms as much as they do to medical devices.

\section{Guardrails for Downstream Risk Estimation}
\label{sec:guardrails}

Validation of an intermediate classifier does not automatically validate the downstream risk, its displayed summary, or its use in a decision. Existing guidance on output postprocessing, clinical evaluation, user understanding, and change control already points in this direction \cite{fda_lifecycle2025,imdrf_gmlp2025,fda_humanfactors2026,fda_transparency2024,fda_pccp2025}. What follows are our proposed task-specific practices, not FDA requirements or a sufficient basis for authorization.

Training to a label is not itself the problem. Minimizing cross-entropy against one-hot labels is a proper scoring rule, and a well-specified model estimates the posterior over states rather than only its most likely value. The problem enters when models are selected by accuracy, which is blind to where the non-modal probability lies; when the reference label was already hardened, as when a reader records only a final category or a claim records only a code; and when the interface exports only the top label. The argument that a model trained for accuracy produces optimal labels is correct for identifying the state, but it does not address whether those labels carry the right outcome risk.

Table~\ref{tab:guardrails} gives six guardrails, each with what a developer produces and what a reviewer checks. One audit quantity deserves mention. For a threshold $\tau$, the disagreement rate
\begin{equation}
D_\tau=P\bigl(\text{the decisions based on } \Rsoft \text{ and } \Rhard \text{ differ}\bigr)
\label{eq:audit-disagreement}
\end{equation}
can be estimated without outcome labels and should be reported separately in each direction. It measures how often compression changes a calculated decision, not how often that decision is wrong; outcome data and a decision model are needed for the latter.

\begin{table}[htbp]
\centering\small
\renewcommand{\arraystretch}{1.2}
\begin{tabularx}{\textwidth}{@{}>{\raggedright\arraybackslash}p{0.22\textwidth}>{\raggedright\arraybackslash}X>{\raggedright\arraybackslash}X@{}}
\toprule
Guardrail & What the developer produces & What a reviewer checks \\
\midrule
1. State what the numbers mean &
A specification of the state partition, reference-label protocol, outcome with its time horizon and care policy, target population, and whether the output is a posterior (under which prevalence) or a likelihood. &
That state probability $P(E\given X)$, state risk $P(A\given E)$, and label risk $P(A\given C)$ are kept distinct; that labels cannot silently become recorded facts; that the classifier and risk table use compatible definitions. \\
2. Select for the distribution, not the label &
Model and calibration-map selection by proper scoring rules (log loss, multiclass Brier score) and full-vector calibration \cite{guo2017,kull2019}, on data disjoint from training. &
Per-class reliability diagnostics; evidence that accuracy or top-class confidence was not the sole selection criterion. \\
3. Validate the downstream risk &
A risk table estimated under the same protocol and population, the label-only comparator $g(C)$, held-out outcome performance of $\Rsoft$, and a test of whether patient data add prognostic information within states. &
Comparison of $\Rsoft$ with $g(C)$, the modal-state lookup, and a direct outcome model where feasible; subgroups; sampling uncertainty; transport to the deployment population. \\
4. Export a justified summary &
An output record carrying the full vector, the propagated scalar, or a partial vector with its omitted mass, never renormalized, plus version identifiers for every component. &
That each reduction (argmax, top-$k$, rounding) preserves what its intended task requires. \\
5. Audit compression against the decision margin &
The distribution of $\Delta$, the disagreement rate $D_\tau$ in each direction, and the fraction of cases the agreement condition settles. &
How inconclusive cases are handled; that failure of a sufficient condition is not read as evidence of error, nor its success as clinical safety. \\
6. Test users and revalidate after changes &
Human-factors evidence that users distinguish state probability from outcome risk, and a change-control and outcome-monitoring plan. &
Comprehension and decisions rather than visual preference; fallback use and delays; repeated validation when the classifier, table, population, or interface changes. \\
\bottomrule
\end{tabularx}
\caption{Proposed guardrails for two-stage risk estimation, for developers, regulatory reviewers, and payers. Evidence should be proportionate to the intended task and its consequences. This is not an FDA checklist or a substitute for device-specific evaluation.}
\label{tab:guardrails}
\end{table}

\section{Discussion and Conclusion}

Accuracy is not risk. A model can select the most likely diagnosis correctly and still lead a clinician, a payer, or a regulator to underestimate the danger to the patient. Avoiding this means answering four separate questions. Does the model's output describe the population where it is used, including its case mix? Are its probabilities calibrated? Does the risk table still apply once the patient's information is known? What does the downstream system keep from the output, and how does it use it? The output specification addresses the first, calibration the second, rigidity the third, and the results of Sections~\ref{sec:hardening} and~\ref{sec:payers} the fourth. Answering one does not answer the others.

The distinction between a predicted and an actual state is especially important for evaluation. The modal-state lookup is not the strongest categorical baseline. A properly estimated label risk $g(C)$ removes the plug-in error but still loses the variation within each label, and the same distinction governs health insurance coverage decisions, where the labels are diagnosis codes and the risk tables are estimated from them. Propagation has limits of its own. It recovers the patient's risk when the within-state risks do not depend on the particular patient, but when they do, the rigidity term in \eqref{eq:total-discrepancy} can make it worse than a well-chosen label even with a correct posterior and an accurate marginal risk table.

The scope of these results should be stated once and plainly. The clinical examples are hypothetical, and the risk tables are illustrative. The agreement conditions are sufficient, not necessary, and agreement between calculations is not evidence of clinical correctness. Mathematical implications, proposed evaluation practices, and demonstrated clinical benefit are distinct levels of evidence; this paper establishes the first and motivates the second. Reference labels can themselves be uncertain \cite{mimori2021}, fitted probabilities and risk tables have sampling error, overlapping diagnoses require a richer state space, and whether clinicians decide better with probabilistic outputs is a question for reader and prospective studies \cite{kompa2021,dratsch2023}.

A predicted label is not an observed state. Training to a label is not the fault; selecting, exporting, and reasoning with only the label is. Jeffrey's rule makes the conditional-risk assumption explicit, stating the prevalence behind an output makes explicit whether it transfers to a new setting, and the results of Section~\ref{sec:hardening} identify what is lost and when decisions must agree. The design principle is to preserve the information the downstream task requires, whether that is a full distribution, a propagated scalar, or another justified summary, and to say what that information means.

\section*{Statements and Declarations}
\noindent\textbf{Funding.} No funding to report.

\smallskip
\noindent\textbf{Competing interests.} No conflicts of interest to report.

\smallskip
\noindent\textbf{Ethical approval.} Not applicable. This work uses no patient data. All clinical examples are hypothetical.

\smallskip
\noindent\textbf{Data and code availability.} Not applicable. This work uses no data, and every calculation is shown in the text and Appendix.

\smallskip
\noindent\textbf{Use of artificial intelligence.} The authors used ChatGPT (OpenAI) and Claude (Anthropic) during manuscript preparation for brainstorming, literature discovery, organization, and language editing. ChatGPT also assisted with mathematical development during revision. All mathematical arguments, calculations, citations, clinical claims, and conclusions remain the responsibility of the authors.

\begingroup
\small
\bibliographystyle{plainurl}
\bibliography{Jeffreys_Rule}
\endgroup

\clearpage
\appendix
\renewcommand{\thesection}{\Alph{section}}
\section*{Appendix}
\addcontentsline{toc}{section}{Appendix}

\section{Notation and Standing Assumptions}
\label{app:notation}

We assume $n\ge2$, $P(E=i)>0$ for the states retained, and the existence of regular conditional probabilities. Conditional statements involving continuously distributed $X$ or $Q$ are understood almost surely. Model predictions and risk tables are treated as fixed when evaluating a target population; sampling uncertainty in fitted quantities is a separate issue. A risk table estimated in another population requires the corresponding transport assumption. The modal index $c(x)$ is the smallest index attaining $\max_i q_i(x)$.

\begin{table}[htbp]
\centering\small
\renewcommand{\arraystretch}{1.15}
\begin{tabularx}{\textwidth}{@{}>{\raggedright\arraybackslash}p{0.19\textwidth}>{\raggedright\arraybackslash}p{0.33\textwidth}>{\raggedright\arraybackslash}X@{}}
\toprule
Symbol & Definition & Plain-language reading \\
\midrule
$P$ & Probability over patient encounters in the specified setting & The population being described \\
$X$, $x$ & Patient information supplied to the model & What the model sees \\
$E$, $n$ & Intermediate state, one of $n$ mutually exclusive states & The diagnosis or imaging category \\
$A$, $Y$ & Outcome event and its indicator & What the patient is at risk of \\
$p_i(x)$ & $P(E=i\given X=x)$ & The correct probability of state $i$ for this patient \\
$q_i(x)$, $Q$ & Model's reported probability of state $i$; $Q=q(X)$ & What the model says \\
$r_i$ & $P(A\given E=i)$ & Risk-table entry for patients actually in state $i$ \\
$c(x)$, $C$ & Index of the largest $q_i(x)$; $C=c(X)$ & The hard label \\
$g(j)$ & $P(A\given C=j)$ & Risk among patients the model labels $j$ \\
$\eta(x)$ & $P(A\given X=x)$ & The patient's true risk given everything the model saw \\
$\Rsoft(x)$ & $\sum_i q_i(x)r_i$ & Propagated risk \\
$\Rhard(x)$ & $r_{c(x)}$ & Modal-state lookup \\
$\Delta(x)$, $B(x)$ & $\Rsoft-\Rhard$ and its bound & What hardening changes, and the most it could change \\
$\tau$, $d_\tau$ & Threshold and the rule $d_\tau(t)=\mathbf 1\{t\ge\tau\}$ & Act when risk is at least $\tau$ \\
$D_\tau$ & Probability that $\Rsoft$ and $\Rhard$ lead to different decisions at $\tau$ & How often hardening changes a decision \\
$\mu_i(x)$, $\delta_i(x)$ & $P(A\given E=i,X=x)$ and $\mu_i(x)-r_i$ & Within-state risk for this patient, and its departure from the table \\
$\epsilon_i(x)$, $e(x)$ & Tolerance on $\delta_i$ and the resulting envelope & How wrong the table could be, and how wrong the propagated risk could be \\
$K$, $m_K$, $a_K$ & Retained classes, omitted mass, retained contribution & What a truncated output keeps and drops \\
\bottomrule
\end{tabularx}
\caption{Notation used in the paper and Appendix.}
\label{tab:notation}
\end{table}

\section{Calibration and Exact Propagation}
\label{app:calibration}

Full-vector calibration means
\begin{equation}
P(E=i\given Q)=Q_i\qquad\text{for every }i.
\label{eq:vector-calibration}
\end{equation}
This concerns groups of patients receiving the same output and does not imply $q_i(x)=p_i(x)$. Calibration of the winning probability or of each coordinate separately does not generally imply \eqref{eq:vector-calibration} \cite{kull2019,vaicenavicius2019}.

\begin{proposition}[Risk conditional on the model output]
\label{prop:output}
Suppose \eqref{eq:vector-calibration} holds, $r_i=P(A\given E=i)$, and $A\indep Q\given E$. Then $\Rsoft(X)=P(A\given Q)$, and the propagated scalar is outcome-calibrated, $P(A\given\Rsoft(X))=\Rsoft(X)$.
\end{proposition}
\begin{proof}
Conditional total probability gives
\[
P(A\given Q)=\sum_i P(A\given E=i,Q)P(E=i\given Q)=\sum_i r_iQ_i.
\]
Conditioning this equality on the scalar $\Rsoft(X)$ proves the second claim.
\end{proof}

\begin{proposition}[Risk conditional on the patient data]
\label{prop:ci}
If $q_i(X)=p_i(X)$ for every $i$, $r_i=P(A\given E=i)$, and $A\indep X\given E$, then $\Rsoft(X)=\eta(X)$.
\end{proposition}
\begin{proof}
By conditional total probability and rigidity, which gives $P(A\given E=i,X)=r_i$,
\[
P(A\given X)=\sum_i P(A\given E=i,X)P(E=i\given X)=\sum_i r_ip_i(X).
\]
\end{proof}

Neither conditional-independence condition, $A\indep Q\given E$ in Proposition~\ref{prop:output} or $A\indep X\given E$ in Proposition~\ref{prop:ci}, is necessary, since weighted within-state discrepancies can cancel, so both results give sufficient conditions rather than characterizations. Calibration alone provides no pointwise guarantee about $\eta(x)$.

\section{Proofs for Section~\ref{sec:hardening}}
\label{app:proofs}

\begin{proof}[Proof of Proposition~\ref{prop:gap}]
Subtract $r_c\sum_iq_i=r_c$ from \eqref{eq:soft-risk}; the $i=c$ term vanishes. The triangle inequality gives the bound, which is attained in the two-state case. If the modal state has the smallest risk, $\Delta\ge0$; if the largest, $\Delta\le0$.
\end{proof}

\begin{proof}[Proof of Corollary~\ref{cor:agreement}]
By Proposition~\ref{prop:gap}, $\Rsoft\in[\Rhard-B,\Rhard+B]$, and the strict inequality places this interval entirely on one side of $\tau$.
\end{proof}

\begin{proof}[Proof of Proposition~\ref{prop:brier}]
Under the hypotheses of Proposition~\ref{prop:output}, $S=\E[Y\given Q]$. Since $C$ is a function of $Q$, the tower property gives $\E[S\given C]=\E[Y\given C]=g(C)$. Also $\E[Y-S\given Q]=0$ and $S-T$ is $Q$-measurable, so expanding the squares gives
$\E[(Y-T)^2]-\E[(Y-S)^2]=\E[(S-T)^2]$.
Write $S-T=(S-g(C))+(g(C)-T)$. The cross term has expectation zero after conditioning on $C$, leaving the two terms of \eqref{eq:brier-decomposition}. Under Proposition~\ref{prop:ci}, $S=\eta(X)$, so the comparison is with the optimal predictor based on all of $X$.
\end{proof}

\section{Partial Outputs and Renormalization}
\label{app:partial}

Fix $x$ and let $K\subsetneq\{1,\ldots,n\}$ be the retained indices. Define
\[
a_K=\sum_{i\in K}q_i r_i,\qquad m_K=1-\sum_{i\in K}q_i,\qquad
r_K^-=\min_{i\notin K}r_i,\qquad r_K^+=\max_{i\notin K}r_i,
\]
the retained contribution to the propagated risk, the omitted mass, and the smallest and largest omitted-state risks.

\begin{proposition}[Omitted-mass bound]
\label{prop:omitted}
The full propagated risk satisfies
\begin{equation}
L_K:=a_K+m_Kr_K^-\ \le\ \Rsoft\ \le\ a_K+m_Kr_K^+=:U_K.
\label{eq:omitted}
\end{equation}
Consequently, $U_K<\tau$ establishes $d_\tau(\Rsoft)=0$, and $L_K\ge\tau$ establishes $d_\tau(\Rsoft)=1$.
\end{proposition}
\begin{proof}
Each omitted term satisfies $q_i r_K^-\le q_i r_i\le q_i r_K^+$. Summing, using $\sum_{i\notin K}q_i=m_K$, and adding $a_K$ gives \eqref{eq:omitted}.
\end{proof}

The bound is pointwise, so $K$ may depend on the reported vector, as in a top-$k$ list. Its width is $m_K(r_K^+-r_K^-)$. If $L_K<\tau\le U_K$, the check is inconclusive; it does not show that any decision is wrong. For the two vectors of Section~\ref{sec:hardening}, retaining only the leading class gives the same interval $[0.013,0.277]$ for both, and their full risks attain opposite endpoints. In the health insurance coverage example of Section~\ref{sec:payers}, $a_K=0.0007$, $m_K=0.30$, $r_K^-=0.10$, and $r_K^+=0.50$, so $[L_K,U_K]=[0.0307,0.1507]$.

Renormalization differs from truncation. When $m_K<1$, replacing the retained probabilities by $q_i/(1-m_K)$ gives $R_{\mathrm{ren},K}=a_K/(1-m_K)$, a state-risk average additionally conditioned on $E\in K$. For $0<m_K<1$, with $\bar r_{K^c}=m_K^{-1}\sum_{i\notin K}q_i r_i$,
\begin{equation}
\Rsoft=(1-m_K)R_{\mathrm{ren},K}+m_K\bar r_{K^c},\qquad
R_{\mathrm{ren},K}-\Rsoft=m_K(R_{\mathrm{ren},K}-\bar r_{K^c}),
\label{eq:renormalization}
\end{equation}
so renormalization can increase or decrease the risk. In the coverage example of Section~\ref{sec:payers}, rescaling among the benign diagnoses gives $R_{\mathrm{ren},K}=0.0007/0.70=0.001$, and rescaling among the serious ones gives $0.055/0.30\approx0.183$.

\section{Sensitivity to Model Error}
\label{app:sensitivity}

Let $\mu_i(x)=P(A\given E=i,X=x)$ and $\delta_i(x)=\mu_i(x)-r_i$, where $r$ is the table actually used, so that departures of the table from target-population risks are absorbed into $\delta_i$. Since $\eta=\sum_ip_i\mu_i$, subtracting \eqref{eq:soft-risk} and adding $\Delta$ gives \eqref{eq:total-discrepancy}.

\begin{proposition}[A sensitivity envelope]
\label{prop:sensitivity}
If $|\delta_i(x)|\le\epsilon_i(x)$ on states with positive $p_i(x)$, then
\begin{equation}
|\eta(x)-\Rsoft(x)|\le e(x):=(r_{\max}-r_{\min})\TV(p(x),q(x))+\sum_i p_i(x)\epsilon_i(x),
\label{eq:sensitivity}
\end{equation}
where $\TV(p,q)=\tfrac12\sum_i|p_i-q_i|$. Consequently, $|\Rhard-\tau|>B+e$ implies agreement of the hard, propagated, and true-risk decisions, and $|\Rsoft-\tau|>e$ implies agreement of the propagated and true-risk decisions.
\end{proposition}
\begin{proof}
The positive and negative parts of $p-q$ each sum to $\TV(p,q)$, so 
\[
|\sum_i(p_i-q_i)r_i|\le(r_{\max}-r_{\min})\TV(p,q).
\]
The triangle inequality gives \eqref{eq:sensitivity}. Combined with Proposition~\ref{prop:gap}, both $\Rsoft$ and $\eta$ lie within $B+e$ of $\Rhard$, and the strict margins keep the relevant risks on one side of $\tau$.
\end{proof}

These are sensitivity statements, not confidence intervals. They require justified bounds on posterior and within-state errors, and calibration supplies neither. For a partial output, any justified bound $|\eta-\Rsoft|\le e$ widens \eqref{eq:omitted} to $\eta\in[\max\{0,L_K-e\},\min\{1,U_K+e\}]$.

In the BI-RADS example with $q_3(x)=0.70$, suppose the reported probabilities are correct and each within-category risk differs from its table value by at most $\epsilon$. Then $e=\epsilon$ and $|\eta-0.025|\le\epsilon$. A tolerance of $0.002$ gives $[2.3\%,2.7\%]$, but a tolerance of $0.005$ gives $[2.0\%,3.0\%]$ and no longer guarantees a risk strictly above 2\%. These are stipulated tolerances, not estimated uncertainty intervals.

\end{document}